\documentclass[12pt,a4paper]{article}

\usepackage{graphicx}

\newcommand{\be}{\begin{equation}}

\newcommand{\ee}{\end{equation}}
\newcommand{\ba}{\begin{eqnarray}}
\newcommand{\ea}{\end{eqnarray}}
\newcommand{\nn}{\nonumber}
\newcommand{\e}{{\rm e}}
\newcommand{\ie}{{\em i.e.\ }}
\newcommand{\eg}{{\em e.g.\ }}

\newcommand{\tr}{{\rm tr}}

\newcommand{\cN}{{\cal N}}

\newcommand{\cW}{{\cal W}}

\renewcommand{\P}{\Phi}

\renewcommand{\a}{\alpha}

\newcommand{\bt}[1]{{\bar t}}

\begin{document}

\begin{titlepage}

\date{}
\vspace{-2.0cm}
\title{
{\vspace*{-6mm} %\small
%{\bf Preliminary draft}\\
%\hfill \parbox{40mm}{
\begin{flushright}
\small{DESY 05-120}\\
\small{LAPTH-1107/05 }\\
\small{ROM2F/2005/12}
\end{flushright}
\vspace{10mm}
%[15mm] }
}%[-12mm]
New results in the deformed ${\cal{N}}=4$  SYM  theory
\vspace*{0mm}}
\author{G.C.\ Rossi$^{a,b}$,\quad E.\ Sokatchev$^{c}$,
\quad Ya.S.\ Stanev$^{a}$ \\[5mm]
%\vskip{.5cm}
  {\small $^a$ Dipartimento di Fisica, Universit\`a di  Roma
   ``{\it Tor Vergata}''} \\
  {\small and INFN, Sezione di Roma 2}\\
  {\small Via della Ricerca Scientifica - 00133 Roma, Italy}\\
  {\small $^b$ John von Neumann-Institut f\"ur Computing  NIC}\\
  {\small Platanenallee 6, D-15738 Zeuthen, Germany}\\
  {\small $^c$  Laboratoire d'Annecy-le-Vieux de
Physique Th\'{e}orique  LAPTH,}\\
  {\small B.P. 110,  F-74941 Annecy-le-Vieux,
  France\footnote{UMR 5108 associ{\'e}e {\`a}
 l'Universit{\'e} de Savoie} }}
\maketitle
\vspace*{0mm}
%%%%%%%%%%%%%%%%%%%%%%%%%%%%%%%%%%%%%%%%%

\abstract
{We investigate various perturbative properties of the deformed $\cN=4$ SYM theory.
We carry out a three-loops calculation of the chiral matter superfield
propagator and derive the condition on the couplings for maintaining
finiteness at this order. We compute the 2-, 3- and 4-point functions
of composite operators of dimension 2 at two loops.
We identify all the scalar operators (chiral and non-chiral) of bare dimension 4
with vanishing one-loop anomalous dimension. We compute some 2- and 3-point
functions of these operators at two loops and argue that the observed
finite corrections cannot be absorbed by a finite renormalization
of the operators.
}

\vspace{2truecm}
%\noindent\small {{\it Keywords:}}
\thispagestyle{empty}
\end{titlepage}

%\vskip .1cm
%\small {$^{*)}${\it Correspondence to:} G.C. Rossi, Phone:
%+39-0672594571; FAX: +39-062025259; E-mail address: rossig@roma2.infn.it}
\vfill
\newpage

\section{Introduction}
\label{sec:INTRO}

Recently there has been a renewed interest~\cite{LS}--\cite{deMelloKoch:2005vq}
in the deformed $\cN=4$ supersymmetric Yang-Mills (SYM)
theory with superpotential\footnote{The trace is over the colour indices
of the fundamental representation of the $SU(N)$ gauge group. The
generators, $T^{a}$, of the fundamental representation are normalized
according to $\tr(T^{a}T^{b})=\frac{1}{2} \delta^{ab}$.}
\be
{\cal W} =  g \ \kappa \int d^4x \ d^{2}\theta \
\tr \left( \Phi^{1} \ [\Phi^{2}, \Phi^{3}]_{\omega} \right) \ + \ {\rm h.c.} \  \ ,
\label{superp}
\ee
where $g$ is the SYM coupling constant and the deformed commutator
is defined as
\be
  [A,B]_{\omega} = \omega \ AB -  BA \, .
\label{defcom}
\ee
The parameter $\kappa$ can be taken real since its phase can
be absorbed into a redefinition of the chiral superfields,
while $\omega$ is in general complex\footnote{We use
a slightly different notation than in~\cite{LuninM,FG}.
The correspondence is given by $\omega=q^2=\e^{2i\pi\beta}$
(with $\beta$ complex) and $g\kappa = h/q$. The reason for our choice is that
in all the formulae only the quantities $\kappa^2$ and
$\omega$ appear.}.

The undeformed $\cN=4$ SYM is recovered when $\kappa =\omega= 1$,
where the action has a manifest $SU(3)\times U_R(1)$ invariance.
For generic values of $\omega$, $\cN=4$ supersymmetry is reduced
to $\cN=1$, and the $SU(3)$ symmetry is broken down to
$U(1)\times U(1)$~\cite{LuninM}. The latter can be chosen to act as follows
\ba
(\P^1,\ \P^2,\ \P^3)&\rightarrow
&(\e^{2i\a_1}\P^1,\ \e^{-i\a_1}\P^2,\ \e^{-i\a_1}\P^3 )\, , \nn \\
(\P^1,\ \P^2,\ \P^3 ) & \rightarrow
&(\P^1,\ \e^{i\a_2}\P^2,\ \e^{-i\a_2}\P^3 )\, .
\label{U1U1}
\ea
The key observation is that apparently many of the interesting properties
of the $\cN=4$ SYM theory are preserved by the deformation. First of all, it is
believed that the deformed theory is finite, provided the couplings satisfy some
condition which ensures the vanishing of all $\beta$-functions. A general
argument to this effect has been given in~\cite{LS}. This claim has been
explicitly verified to order $g^4$ in~\cite{FG,PSZ}, using results
of~\cite{LS,JackJN,GRS}. In our notation, the condition reads
\be
\kappa^2={2\ N^2\over(N^2-2)(\omega\bar\omega+1)+2(\omega+\bar\omega)}\, .
\label{1Lcondc}
\ee
Another feature retained by the deformed theory is the existence of
composite operators with vanishing anomalous dimension.
A list of such single trace chiral primary operators (CPO) has been
proposed in~\cite{BJL}.
These include $\tr(\P^1\P^1)$,  $\tr(\P^1\P^1\P^1)$, etc.,
which in the undeformed theory belong to short ($1/2$ BPS) supermultiplets.
In~\cite{FG} it has been shown that also CPOs of dimension
two of the form $\tr(\Phi^1 \Phi^2)$ have vanishing order $g^2$ anomalous dimension.
This is non-trivial since, while in the undeformed $\cN=4$ theory
$\tr(\Phi^1 \Phi^2)$ and  $\tr(\P^1\P^1)$ belong to the same supermultiplet,
in the deformed theory the two operators are not related by supersymmetry.
In~\cite{PSZ} this result has been extended to order $g^4$.
Neither multi-trace, nor non-chiral operators have
been considered in the literature.

In this paper we investigate further the non-renormalization
properties of the deformed theory. To simplify the calculations,
we shall use the obvious observation that if some quantity is known in
$\cN=4$ SYM, then it is sufficient to compute the difference
between the  perturbative  corrections in the deformed and undeformed cases.

The outline of the paper is as follows.  In Section~\ref{propagatorg6} we
show that a new condition is necessary for the finiteness of the
deformed theory at three loops (order $g^6$). We give the explicit
form of this condition and discuss its general solution.
Section~\ref{composites} is devoted to a systematic search for all operators
of dimension $\Delta \leq 4$ with vanishing order $g^2$ anomalous dimensions.
Among them we find non-chiral operators, both of dimension
2 and of dimension 4, as well as numerous mixtures of single and double trace
operators of dimension 4. We also compute the perturbative  corrections to some
2-, 3- and 4-point functions at order $g^4$ and argue that the observed
finite corrections cannot be absorbed by any finite renormalization
of the operators. Some conclusions can be found
in Section~\ref{Conclusions}.

\section{Perturbative corrections to the correlation functions}
\label{propagatorg6}

We shall write the action of the deformed theory in the form
\be
S_{\kappa} = S_0 + S_v + \cW_{\kappa} \, ,
\label{Sh}
\ee
where $S_0$ contains the kinetic terms, and
$S_v$ is the part of the standard $\cN=4$ SYM action involving the
couplings of the gauge superfield, $V$.
Finally $\cW_{\kappa}$ is the deformed superpotential\footnote{
We shall use the shorthand notation $\cW_{\kappa}$ instead of
$\cW_{\kappa,\omega}$, keeping in mind
that the superpotential also depends on $\omega$.}
\ba
\cW_{\kappa} &=& g \ \kappa \ \int d^4x d^2\theta \ \left( \ \omega \
\tr  (\Phi^1 \Phi^2 \Phi^3 )
-  \ \tr ( \Phi^1 \Phi^3 \Phi^2 )  \right)  \ + \nn \\
&+& g\ \kappa \ \int d^4x d^2 \bar \theta \ \left( \  \bar \omega \
\tr (\Phi_1^{\dagger} \Phi_3^{\dagger}
\Phi_2^{\dagger} )
-  \ \tr ( \Phi_1^{\dagger} \Phi_2^{\dagger}
\Phi_3^{\dagger} ) \right)\, .
\label{Wh}
\ea
As a matter of principle, both parameters, $\kappa$ and $\omega$, could
depend on $g$ and the number of colours, $N$. We shall assume, however,
that they have a Taylor series expansion in powers of $g$ around $g=0$.

In our notations the action of the undeformed $\cN=4$ SYM theory reads
($\kappa=\omega= 1$)
\be
S_g = S_0 + S_v + \cW_g \, ,
\label{Sg}
\ee
with $\cW_g$ the $\cN=4$  superpotential
\be
\cW_g = g \ \int d^4x d^2\theta  \ \tr \left(\Phi^1 \left[\Phi^2 ,
\Phi^3 \right] \right)
- g \ \int d^4x d^2 \bar \theta  \ \tr \left( \Phi_1^{\dagger}
\left[ \Phi_2^{\dagger} ,
\Phi_3^{\dagger}\right] \right)
\, .
\label{Wg}
\ee

\subsection{General considerations}

We want to compute the order $g^{2n}$  correction  in the deformed theory
to a given $\ell$-point correlation function,
$\langle {\cal O}_1(x_1) \dots {\cal O}_{\ell}(x_{\ell})\rangle$,
where ${\cal O}_i(x_i)$ are some local (fundamental or composite)
operators. To this end we have to evaluate the correlator
\be
G_{\kappa}^{2n}(x_1, \dots ,x_{\ell}) =\langle\left.
\e^{ S_v+ \cW_{\kappa} }\right|_{g^{2n}}
{\cal O}_1(x_1)  \dots {\cal O}_{\ell}(x_{\ell})\rangle \, ,
\label{Ghn}
\ee
where by $\e^{S_v+\cW_{\kappa}}|_{g^{2n}}$ we denote all terms of order
$g^{2n}$ in the expansion of the exponent. Since both $S_v$ and $\cW_{\kappa}$
are non-linear in $g$, one gets in general a rather complicated expression.

The same computation in $\cN=4$ SYM will give
\be
G_g^{2n} (x_1, \dots ,x_{\ell}) =\langle\left. \e^{S_v+\cW_g}\right|_{g^{2n}}
{\cal O}_1(x_1)  \dots {\cal O}_{\ell}(x_{\ell})\rangle\, .
\label{Ggn}
\ee
Both computations are very involved as soon as $n$ gets large ($n\geq 2$),
since there are numerous diagrams which contribute.
Most of them, however, come from $S_v$ which contains all the
non-polynomial interactions involving the gauge field. This suggests
that it might be simpler to compute the difference
\be
\delta G^{2n} (x_1, \dots ,x_{\ell}) =
G_{\kappa}^{2n} (x_1, \dots ,x_{\ell}) - G_g^{2n} (x_1, \dots ,x_{\ell}) \, ,
\label{deltaG_1}
\ee
between the corrections to the correlator in the two theories.

In particular, if we are interested in the corrections
 to the 2-point function of an operator, say ${\cal O}$, which is known to
be protected in $\cN=4$ SYM (\ie for which $G_g^{2n}=0$, $n\geq 1$),
then the vanishing of $\delta G^{2n}$ implies in an obvious way
$G_{\kappa}^{2n}=0$. Hence, if $\delta G^{2n}=0$, the operator ${\cal O}$
will have vanishing anomalous dimension at order $g^{2n}$ also in the
deformed theory.

Inserting eqs.~(\ref{Ghn}) and~(\ref{Ggn}) in~(\ref{deltaG_1}), one obtains
\be
\delta G^{2n} (x_1, \dots ,x_{\ell})= \langle \left\{
\left. \e^{ S_v+ \cW_{\kappa} } \right|_{g^{2n}} -
\left. \e^{ S_v+ \cW_g } \right|_{g^{2n}} \right\}
{\cal O}_1(x_1)  \dots {\cal O}_{\ell}(x_{\ell})\rangle\, .
\label{deltaG_2}
\ee
It is useful to rewrite the factor in the braces as
\ba
 \left. \e^{ S_v+ \cW_{\kappa} } \right|_{g^{2n}} -
\left. \e^{ S_v+ \cW_g } \right|_{g^{2n}}
&=& \sum_{p=1}^{2n-1} \left. e^{ S_v} \right|_{g^{2n-p}}
\left(\left. e^{ \cW_{\kappa} } \right|_{g^p} -
\left. e^{ \cW_g } \right|_{g^p} \right)  \ + \nn \\
&+& \left(\left. e^{ \cW_{\kappa} } \right|_{g^{2n}} -
\left. e^{ \cW_g } \right|_{g^{2n}} \right)
\, .
\label{expWh-Wg}
\ea
As expected, the most complicated term corresponding
to $p$=0 has disappeared from  the right hand side (r.h.s.). For
future convenience we have separated from the rest
the pure superpotential interaction. This equations can
be used to simplify the computation of all correlation
functions whose expressions in $\cN=4$ SYM are explicitly known.

\subsection{Correction to the propagators at order $g^6$}

Let us start by   considering the corrections to the chiral propagator
\be
\langle
\Phi^1_{a}(x_1, \theta_1) \Phi_{1 \ b}^{\dagger}(x_2,\bar\theta_2) \rangle
\label{Cprop}
\ee
in the deformed theory. As we already recalled, the vanishing of the
coefficient to the (logarithmically divergent) order $g^2$ contribution
requires the relation~(\ref{1Lcondc}) between $\kappa$ and $\omega$ to
hold. Then, if $\kappa$ and $\omega$ are assumed not to depend on $g$,
the finiteness\footnote{The difference between the order $g^4$ corrections in the
deformed and undeformed theories is zero. The finite correction in $\cN=4$
SYM has been computed in ref.~\cite{GRS}.} of the order $g^4$ correction
follows without further constraints~\cite{JackJN} (for an explicit check
see also~\cite{PSZ}). If one allows a $g$-dependence in $\kappa$
and/or $\omega$, the finiteness of the order $g^4$ correction
to the propagator implies the absence of order $g^2$ corrections
to eq.~(\ref{1Lcondc}). This means that the solution to the
condition for the finiteness of the  chiral propagator
(\ie vanishing of $\gamma_\Phi$) at orders $g^2$ and $g^4$
can be written in the more general form
\be
\kappa^2 = {2 \  N^2 \over (N^2-2)(\omega\bar\omega+1)+2(\omega+\bar\omega)}
 + {\rm O} (g^4) \, .
\label{1Lcond}
\ee

Let us now consider the order $g^6$ corrections to the chiral
propagator~(\ref{Cprop}). As explained before, to this end we
shall make use of eqs.~(\ref{deltaG_2}) and~(\ref{expWh-Wg}),
and compute only the difference of the order $g^6$ corrections
to the propagator in the deformed and in the $\cN=4$ SYM theory. At first
sight this still looks like a very difficult task, since there are very many
different superdiagrams. The crucial and rather surprising observation,
however, is that through order $g^6$, if $\kappa$ and $\omega$ are related
by eq.~(\ref{1Lcond}) then  all contributions coming from
the first line in the r.h.s.\ of eq.~(\ref{expWh-Wg}) have a vanishing colour
factor\footnote{To compute the complicated colour traces we use a Maple
program to implement the split/join rules for $SU(N)$~\cite{Cvit}.}.
Let us stress that the unspecified ${\rm O} (g^4)$ correction in
eq.~(\ref{1Lcond}) does not modify this conclusion.

As a consequence of this analysis, we conclude that at order $g^6$ we will
only have to compute the correction to the chiral propagator in the deformed
theory of the form
\be
\langle \left. ( e^{ \cW_{\kappa} }  -
 e^{ \cW_g } )\right|_{g^{6}}
\Phi^1_{a}(x_1, \theta_1) \Phi_{1 \ b}^{\dagger}
(x_2, \bar \theta_2)\rangle \, .
\label{prop_Wh-Wg}
\ee
This is a drastic simplification, because one is
effectively left with only the task of computing the propagator
corrections coming exclusively from the superpotential.
Moreover, since we already cancelled the propagator
corrections at lower orders, we have to
consider only primitive divergent superdiagrams (\ie those which
do not contain divergent subdiagrams).
%%%%%%%%%%%%%%%%%%%%%%%
\begin{figure} [ht]
%[!htb]
% [!htbp]
 %\begin{minipage}[t]{\linewidth}
    \centering
    \includegraphics[width=1\linewidth]{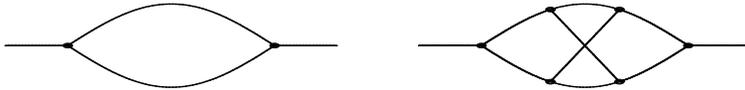}
       \vskip -200pt
  %   \vskip -8cm
\caption{The 1-loop and the 3-loop diagrams.}
  \label{diagr1}
% \end{minipage} \\ [2.5cm]
\end{figure}
There are only two such contributions to order $g^6$ (see figure),
both logarithmically divergent~\cite{AG}. One is given by the one-loop
diagram, because at order $g^6$ it will appear multiplied by
the possibly non-vanishing O($g^4$) term in eq.~(\ref{1Lcond}). The second
is a genuine three-loop nonplanar diagram, which is present
only in the deformed theory, since the corresponding colour
factor in $\cN=4$ SYM is zero. We note that, owing to the structure
of the superpotential, there are no primitive divergent two-loop superdiagrams.
After computing the colour factor of the three-loop diagram, we find
that its contribution is proportional to
\be
g^6\kappa^6\ (\omega-1)\ (\bar \omega -1)\ P(\omega, N)\ {(N^2-4)\over N^3} \, ,
 \label{D3coef}
\ee
where
\be
P(\omega, N) =  ( (\omega^2+\omega+1) ({\bar \omega}^2+\bar \omega+1) - 9 \
\omega \bar \omega  ) \ N^2
 + 5 \ (\omega-1)^2 \ (\bar \omega-1)^2 \, .
\label{Pomega}
\ee
In view of this result we may argue that there are three essentially
different ways to achieve perturbative finiteness at three-loops
(\ie at order $g^6$) if we allow $\kappa$ or $\omega$ to depend on $g$.

1) One may choose the ${\rm O} (g^4)$ term in eq.~(\ref{1Lcond})
so as to compensate the contribution of the three-loop diagram.

2) One can make the coefficient
in~(\ref{D3coef}) proportional to $g^8$, shifting in this way
the problem to the next order. This in turn can be done
either by choosing $\omega= 1+ g \ \omega_1 + \dots $, or by seeking
for an $\omega$ which solves the equation $P(\omega, N)={\rm O}(g^2)$.

3) One may insist that $\kappa$  and $\omega$ do not depend on $g$.
In this case
$\omega$ has to satisfy the equation
\be
( (\omega^2+\omega+1) ({\bar \omega}^2+\bar \omega+1) - 9 \
\omega \bar \omega  ) \ N^2
 + 5\ (\omega-1)^2 \ (\bar \omega-1)^2  = 0 \, .
\label{eq3L}
\ee

As for the gauge field propagator, we find that
eq.~(\ref{1Lcond}) alone is sufficient for its finiteness
both at order $g^4$ and order $g^6$, as
expected on the basis of general
renormalization arguments~\cite{JackJN}.

To summarize, we have proven by an explicit computation that
the deformed theory can be made as finite as
$\cN=4$ SYM also at order $g^6$, and in various ways.
A preliminary investigation shows that the different
scenarios have rather different
implications at the next perturbative
order\footnote{We will not pursue this question any further
in this paper~\cite{NEXT}.}.
We end this Section by noting that,
since the only relevant contribution at order $g^6$ is nonplanar,
as far as the planar limit is concerned, eq.~(\ref{1Lcond}) alone
is sufficient for finiteness also at order $g^6$.

\section{Composite operators with vanishing\\ anomalous dimension}
\label{composites}

In this Section  we present some results for the 2-, 3- and 4-point
functions of the composite operators of naive
dimension $\Delta_0 \leq 4$ in the deformed theory
at orders $g^2$ and $g^4$. We shall assume that
$\kappa$ and $\omega$ satisfy eq.~(\ref{1Lcond}), hence
both the chiral and the gauge field propagator are
finite up to order $g^4$. In practice this also means that there is only
one parameter left in all the formulae, namely $\omega$.

To label operators we shall sometimes use a shorthand notation where
\be
\tr(121\bar3)\ \ {\rm denotes}\ \ \tr(\Phi^1\Phi^2\Phi^1\Phi_3^{\dagger})
\label{notation}
\ee
with all fields at the same point and in the indicated order under the colour
trace. We list operators up to relabelling of the flavour indices.
Hence $(1112)$ is also meant to denote $(1113)$, $(2221)$, etc.
As usual, we shall expand the anomalous dimension of the operator
${\cal O}$ in power series of $g^2$ by writing
\be
\gamma_{\cal O} (g^2) = g^2 \ \gamma^{(1)}_{\cal O} + g^4 \
\gamma^{(2)}_{\cal O} + \dots \, .
\label{gammageneral}
\ee

\subsection{Operators of $\Delta_0=2$ up to order $g^4$}

The 21 different scalar operators of naive dimension $\Delta_0=2$ can be
organized as follows. There are 6 chiral and 6 antichiral operators, namely
\be
{\cal C}^{IJ} = \tr(\Phi^I \Phi^J)  \, , \quad
{\cal C}_{IJ}^{\dagger} = \tr(\Phi_I^{\dagger}\Phi_J^{\dagger})\, ,
\label{d2chiral}
\ee
there are 3 non-chiral, mutually orthogonal, flavour singlet operators
which we shall choose as
\ba
{\cal V}_X &=& 2 \  \tr(\Phi^1 \Phi_1^{\dagger})-\tr(\Phi^2 \Phi_2^{\dagger})-
\tr(\Phi^3 \Phi_3^{\dagger})  \, , \nn \\
{\cal V}_Y &=&  \tr(\Phi^2 \Phi_2^{\dagger})-
\tr(\Phi^3 \Phi_3^{\dagger})  \, , \nn \\
{\cal K}_1 &=& \tr(\Phi^1 \Phi_1^{\dagger})+\tr(\Phi^2 \Phi_2^{\dagger})+
\tr(\Phi^3 \Phi_3^{\dagger}) \, ,
\label{d2fsinglet}
\ea
and 6 operators which are neither chiral nor flavour singlets
\be
{\cal V}^I_J = \tr(\Phi^I \Phi_J^{\dagger}) \quad {\rm for} \ \ I \neq J \, .
\label{d2VIJ}
\ee

In the undeformed $\cN=4$ SYM theory the 20 operators
${\cal C}^{IJ}$, ${\cal C}_{IJ}^{\dagger}$, ${\cal V}^I_J$,
${\cal V}_X$ and ${\cal V}_Y$ are the lowest components
of the short stress-tensor supermultiplet, and hence all have
protected dimension, while ${\cal K}_1$ is the lowest
component of the long Konishi supermultiplet and acquires
anomalous dimension.
The chiral primary operators (CPO) ${\cal C}^{IJ}$ have been already
considered in the literature.
In~\cite{FG} it was shown that their 2-point functions get neither infinite
nor finite corrections at order $g^2$. The absence of anomalous dimension
for these operators was confirmed at order $g^4$ in~\cite{PSZ},
but at this order finite corrections appear.
Our calculations agree with these references. We present here new results
about the non-chiral operators~(\ref{d2fsinglet}) and~(\ref{d2VIJ}).

Since all the 2-, 3- and 4-point functions of all the 21 scalar
operators have already been computed in $\cN=4$ SYM to order $g^4$,
we shall compute only the difference of these correlators between
the deformed and the undeformed theory.
To this purpose the splitting in eq.~(\ref{expWh-Wg}) again
turns out to be very useful.

Our results can be summarized as follows. If the deformation
parameters $\kappa$ and $\omega$ satisfy eq.~(\ref{1Lcond}), then

{\bf -} The 6 operators ${\cal V}^I_J$  defined in  eq.~(\ref{d2VIJ})
have non-vanishing order $g^2$ anomalous dimension given by
\be
\gamma^{(1)}_{{\cal V}^I_J} \ = \ {  N \over (4\pi^2)} \,
{(N^2-4)(\omega-1) (\bar \omega-1) \over
(N^2-2)(\omega\bar\omega+1)+2(\omega+\bar\omega)} \, .
\label{gamma1VIJ}
\ee

{\bf -} The flavour singlets ${\cal V}_X$ and ${\cal V}_Y$, defined in
eq.~(\ref{d2fsinglet}), have vanishing anomalous dimensions both at order
$g^2$ and $g^4$, \ie
\be
\gamma^{(1)}_{{\cal V}_{X,Y}}=0\, ,\qquad\gamma^{(2)}_{{\cal V}_{X,Y}} = 0 \, .
\label{gammaV}
\ee
Their 2- and 3-point functions do not receive any (not even finite)
corrections up to order $g^4$. In fact, since these operators are
the lowest components of the supermultiplets which contain also the
generators of the two $U(1)$ symmetries (see eq.~(\ref{U1U1})),
we expect them to be protected to all orders.

{\bf -} The order $g^2$ and $g^4$ anomalous dimensions of the Konishi scalar
${\cal K}_1$ are the same as in the undeformed $\cN=4$ SYM, \ie
\be
\left. \gamma^{(1,2)}_{{\cal K}_1}\right|_{\rm Deformed} =
\left. \gamma^{(1,2)}_{{\cal K}_1} \right|_{\rm {\cN}=4 \ SYM} \, .
\label{gammaK}
\ee

{\bf -} All the 3- and 4-point functions of the flavour singlets
${\cal O}_{\rm fs} = \{{\cal V}_X$, ${\cal V}_Y, {\cal K}_1 \}$ in any
combination at orders $g^2$ and $g^4$ are {\it exactly equal}
to the corresponding functions in $\cN=4$ SYM. This implies that
up to order $g^4$ the same property will hold for all the $n$-point
functions involving these operators. In formulae
\be
\left. \langle {\cal O}_{\rm fs} (x_1) \dots {\cal O}_{\rm fs} (x_n)
\rangle \right|_{{\rm Deformed}\  g^2, g^4} =
\left. \langle {\cal O}_{\rm fs} (x_1) \dots {\cal O}_{\rm fs} (x_n)
\rangle \right|_{{\rm {\cN}=4}\  g^2, g^4} \, .
\label{Ofs}
\ee
Let us note that this result, combined with the absence of protected
dimension $\Delta_0=4$ scalar operators in the flavour singlet sector
(see below), poses severe constraints on the conformal
Operator Product Expansion (OPE) interpretation of these correlators.

{\bf -} The 3-point functions
$\langle {\cal C}^{IJ} (x_1) \, {\cal C}_{IJ}^{\dagger}(x_2) \,
{\cal V}_{X,Y} (x_3)\rangle $ receive
finite non vanishing corrections at order $g^4$ (but not at order $g^2$).
These corrections affect only the normalizations of the functions and
are proportional to the corrections of the 2-point functions
$\langle {\cal C}^{IJ} (x_1) \, {\cal C}_{IJ}^{\dagger}(x_2)\rangle$
at the same order (see eq.~(\ref{2ptg4}) below).
In fact, since the operators ${\cal V}_{X}$ and ${\cal V}_{Y}$ are
the lowest components of supermultiplets containing conserved currents,
the above 3-point and 2-point functions are related by the Ward identities.

\subsection{Operators of $\Delta_0=3$ up to order $g^4$}

All the  operators of $\Delta_0=3$ with vanishing order $g^2$ anomalous dimension
are chiral (or antichiral) and were all found in~\cite{FG}.
Among them we find that (in the notation introduced in~(\ref{notation}))
\be
{\cal O}^{111} = \tr(111)
\label{d3111}
\ee
has also vanishing order $g^4$ anomalous dimension
\be
\gamma^{(2)}_{{\cal O}^{111}} = 0 \, ,
\label{gammaO111}
\ee
while its 2-point function receives finite corrections at the same order.

The only other operator with vanishing order $g^2$ anomalous dimension is
\be
{\cal O}^{123} = \tr(123)+{(N^2-2)\ \bar \omega +2
\over N^2-2 +2 \ \bar \omega} \ \tr(132) \, .
\ee
This is the first example (we shall give more below) of an
operator with coefficients explicitly depending on the
parameter $\omega$. In this case we could not push
our calculation to the next order, because ${\cal O}^{123}$
(for $\omega\neq 1$) is not protected in the undeformed theory.
As a consequence, it is not sufficient to compute
the difference of the perturbative corrections in the two theories.

\subsection{Operators of $\Delta_0 = 4 $ at order $g^2$ and $g^4$}

We shall consider operators made of scalars only, since at order $g^2$
they form a closed subspace. It should be noted that some of them
(\eg the non-chiral operators listed below) can have
a fermionic contribution. However, since this piece will enter
multiplied by an overall $g$ factor, it will not affect
the lowest-order logarithmic behaviour of the 2-point functions. Thus, the
fermionic contribution is irrelevant as far as only the order $g^2$ anomalous
dimension of the scalar operators is considered and we will omit it.

To compute the anomalous dimensions of the operators at order $g^2$
we use the following (standard) procedure (for the details see
\eg\cite{surprises}).

- We write down all scalar operators ${\cal O}_i$ with given flavour
and bare dimension $\Delta_0$.

- We compute the matrix of the tree-level 2-point functions
\be
F^{(0)}_{ij}=\langle{\cal O}_i(x_1){\cal O}_j^{\dagger}(x_2)
\rangle_{{\rm tree}}\, ,
\label{Tree2pt}
\ee
and notice that the coordinate dependence in all the functions is the same,
\ie $(x_{12}^2)^{-\Delta_0}$.

- We compute at order $g^2$ the matrix of the 2-point functions in the
deformed theory
\be
F^{(1)}_{ij}=\langle {\cal O}_i(x_1){\cal O}_j^{\dagger}(x_2)\rangle_{g^2}\, .
\label{OneL2pt}
\ee
To this order the only relation one needs is eq.~(\ref{1Lcond}),
by which we can express $\kappa$ in terms of $\omega$ and $\bar\omega$.
In this way $F^{(1)}$ will depend only on $N$, $\omega$ and $\bar\omega$.
The coordinate dependence of the logarithmically divergent one-loop
contributions is identical for all diagrams and it is given by
$(x_{12}^2)^{- \Delta_0} \times B(\epsilon,x_{12})$,
where $B(\epsilon,x_{12})$ is the (regularized)
massless box integral. It is the latter that
contains the logarithmic divergence
responsible for the anomalous dimension of the operators.
Hence in both $F^{(0)}$ and $F^{(1)}$
we can factorize out the common coordinate dependence, and consider them
as matrices with numerical entries.

 - The anomalous dimensions of the operators are finally given
by the eigenvalues of the matrix
\be
\left(F^{(0)}\right)^{-1} F^{(1)}\, .
\label{F0F1}
\ee
In particular operators with vanishing  order $g^2$ anomalous dimension
correspond to the zero eigenvalues of this matrix.
Their explicit tree level form in the ${\cal O}_i$ basis is given by the
corresponding eigenvectors. In other words, to find all the operators
of vanishing order $g^2$ anomalous dimension, it is sufficient to find the
kernel of the matrix~(\ref{F0F1}).

We now list all the scalar operators of $\Delta_0=4$ which have
vanishing anomalous dimensions at order $g^2$.
The list is exhaustive for generic $\omega$.
For special values, in particular for $\omega=-1$, there
may be more operators (see \eg eq.~(3.29) in~\cite{FG}).
If some flavour choice is not in the list, then there are no operators with
vanishing order $g^2$ anomalous dimension with that flavour.
In particular, we found no operators with vanishing order $g^2$ anomalous
dimension among the 39 flavour singlets of the type $(I\bar I J \bar J)$.
In order to maximally simplify the coefficients of the various linear
combinations we make reference to some suitably chosen (possibly
non-orthogonal) basis of operators belonging to the kernel of the
matrix~(\ref{F0F1}).

\subsubsection*{Chiral operators}
   $ $

{\bf - flavour} ${1111}$ - There are 2 operators, one single and one double
trace
\be
 \tr(1111) \quad {\rm and }  \quad \tr(11)\,  \tr(11)\, .
\label{d41111}
\ee
Both have vanishing order $g^2$ and $g^4$ anomalous dimensions, \ie
\be
\gamma^{(1)}_{{\cal O}^{1111}} = \gamma^{(2)}_{{\cal O}^{1111}} = 0 \, .
\label{gammaO1111}
\ee
Their 2-point functions receive finite non-vanishing corrections
at order $g^4$.

{\bf - flavour} ${1112}$ - There are 2 operators, one single and one
double trace
\be
 \tr(1112) \quad {\rm and }  \quad \tr(11) \, \tr(12)\, .
\ee
Only one of them, namely
\be
 {\cal O}^{1112} = \tr(1112) -{(N^2-6) \over 2 N} \  \tr(11) \, \tr(12)
\label{O1112}
\ee
has vanishing order $g^2$ and $g^4$ anomalous dimensions, \ie
\be
\gamma^{(1)}_{{\cal O}^{1112}} = \gamma^{(2)}_{{\cal O}^{1112}} = 0 \, .
\label{gammaO1112}
\ee
Its 2-point function receives finite non-vanishing corrections
at order $g^4$.

{\bf - flavour} ${1122}$ - There are 4 operators, two single and two
double trace operators
\be
 \tr(1122) \ , \ \tr(1212) \ , \ \tr(11) \, \tr(22) \quad
{\rm and }  \quad \tr(12) \, \tr(12) \, .
\ee
Two of them have  vanishing order $g^2$ anomalous dimensions.
Both are linear combinations of single and double trace. They can be
taken as the following combinations
\be
2 \ \tr(1122)+\tr(1212)-{(N^2-6)
\over 2 N}(2 \ \tr(12) \, \tr(12)+\tr(11) \, \tr(22) )\, ,
\label{1122o1}
\ee
and
\be
\tr(1122)-\tr(1212) -{N \over 4} \ \tr(11) \ \tr(22)+N \ \tr(12) \, \tr(12)\, .
\ee

{\bf - flavour} ${1123}$ - There are 5 operators, three single and
two double trace operators
\ba
 && \tr(1123) \,  , \ \tr(1132)  \, ,  \ \tr(1213) \, , \nn \\
 && \tr(11) \,\tr(23) \quad {\rm and }  \quad \tr(12) \, \tr(13)\, .
\ea
Two of them have vanishing order $g^2$ anomalous dimensions.
Both are linear combinations of single and double traces. They
can be taken to be
\ba
&& \tr(1123) + \tr(1132)+\tr(1213) \, + \nn \\
&& \quad -{(N^2-6)\over 2 N}
(2 \ \tr(12) \, \tr(13)+\tr(11) \, \tr(23) )\, ,
\label{1123o1}
\ea
and
\ba
&&\tr(1123)-\tr(1132)+{((\bar \omega^2+1)(N^2-2) +4\bar \omega)
\over N^2 (\bar \omega^2-1)} \ \tr(1213) \, + \nn \\
&&-{(N^2-2)((\bar \omega^2+1)(N^2-2)+4\bar \omega)
\over 2 N^3(\bar \omega^2-1)} \ \tr(11) \, \tr(23)  \, +  \\
&&  - {N^4 (\bar \omega+1)^2 -2(\bar \omega^2+1)N^2 +4(\bar \omega-1)^2
\over  N^3(\bar \omega^2-1)}\ \tr(12) \, \tr(13)  \, . \nn
\ea
Let us note the similarity of the mixing coefficients in the
operators~(\ref{O1112}), (\ref{1122o1}) and~(\ref{1123o1}).

\subsubsection*{Non-chiral operators}
  $ $

{\bf - flavour} ${11J \bar J}$ - There are 12 operators, seven
single and five double trace operators. They are
\ba
 && \tr(111 \bar 1)  \ , \ \tr(11) \,  \tr(1\bar 1) \ ,    \nn \\
 &&  \tr(112 \bar 2)  \ ,  \ \tr(11\bar 2 2)  \ ,  \ \tr(121 \bar 2) \  , \
  \tr(11)  \, \tr(2\bar 2) \ ,  \ \tr(12)  \, \tr(1\bar 2) \ ,  \nn \\
  &&  \tr(113 \bar 3)  \ ,  \ \tr(11\bar 3 3)  \ ,  \ \tr(131 \bar 3) \  , \
  \tr(11) \, \tr(3\bar 3) \ ,  \ \tr(13) \, \tr(1\bar 3)\, .
\ea
Two operators have vanishing order $g^2$ anomalous dimensions.
The first one is pure double trace
\be
\tr(11)\left(2 \ \tr(1\bar 1)-\tr(2\bar 2) - \tr(3\bar 3) \right)
= {\cal C} ^{11} \ {\cal V}_X\,
\ee
(see eqs.~(\ref{d2chiral}) and~(\ref{d2fsinglet}) for the notation).
Let us stress that, surprisingly, the operator ${\cal C} ^{11} \ {\cal V}_Y$,
although very similar in structure to ${\cal C} ^{11} \ {\cal V}_X$
has a non-vanishing order $g^2$ anomalous dimension.

The second one can be taken as the following linear combination of single and
double trace terms
\ba
&& 2(N^2-3) \ (\tr(112 \bar 2)+\tr(11\bar 2 2) ) + 12 \ \tr(121 \bar 2) \, + \nn \\
&&-N (N^2-7) \ \tr(11) \, \tr(2\bar 2)-4 N \ \tr(12) \,  \tr(1\bar 2) \, +  \nn \\
&& - (2 \leftrightarrow 3 \ , \ \bar 2 \leftrightarrow \bar 3)\, .
\ea

{\bf - flavour} ${12J \bar J}$: There are 19 operators, twelve single
and seven double trace ones
\ba
 && \tr(123 \bar 3)  \ ,  \ \tr(12\bar 3 3)  \ ,  \ \tr(132 \bar 3)  \ ,
 \ \tr(13 \bar 3 2) \  ,  \ \tr(1 \bar 3 23) \ ,  \ \tr(1\bar 3 3 2) , \nn \\
&&\tr(12)\, \tr(3\bar 3)\ , \ \tr(13)\, \tr(2\bar 3)\ ,\ \tr(1\bar 3)\, \tr(23)
\ , \nn \\
&& \tr(112 \bar 1)  \ , \ \tr(11 \bar 1 2)  \ , \ \tr(121 \bar 1)  \ , \
   \tr(11) \, \tr(2\bar 1) \ ,  \ \tr(12) \, \tr(1\bar 1) \ ,  \nn \\
&& \tr(221 \bar 2)  \ ,  \ \tr(22\bar 2 1)  \ ,  \ \tr(212 \bar 2) \ , \
\tr(22) \, \tr(1\bar 2) \ ,  \ \tr(21) \,  \tr(2\bar 2)\, .
\ea
Again there are two operators with vanishing order $g^2$ anomalous dimensions,
both are linear combinations of single and double traces.
The first one can be taken as
\ba
 && \tr(123 \bar 3)  + \tr(1\bar 3 3 2)
+ N \tr(13) \, \tr(2\bar 3) \, +  \nn \\
&-& \tr(112 \bar 1) - \tr(11 \bar 1 2)  -(N^2-2) \tr(121 \bar 1)
  +N(N^2-3) \tr(12) \, \tr(1\bar 1)  \, +  \nn \\
  &-& (1 \leftrightarrow 2  \ , \ \bar 1 \leftrightarrow \bar 2)\, ,
\ea
while the second has rather complicated mixing coefficients depending on both
$\omega$ and $\bar \omega$. One finds
\ba
&& N^2(N^2\bar \omega(3\omega \bar \omega+1)-2(\bar \omega-1)(4\omega \bar \omega
-3\bar \omega+ \omega-2)) \, \times \nn \\
&&\times (-\omega \tr(123 \bar 3)- \omega \tr(12\bar 3 3) +2 \tr(132 \bar 3)
 - \tr(11 \bar 1 2)  - \tr(221 \bar 2)) \, + \nn \\
&+& N^2(N^2(\omega \bar \omega+3) -2(\bar \omega-1)(2 \bar \omega \omega-\bar \omega+
3 \omega-4)) \, \times  \nn \\
&& \times (- \tr(13 \bar 3 2) +2 \omega \tr(1 \bar 3 23) - \tr(1\bar 3 3 2)
 -  \omega \tr(112 \bar 1)  -  \omega \tr(22\bar 2 1) )  \, + \nn \\
&+& N( \omega-1)((3 \omega \bar \omega^2-  \bar \omega \omega+
\bar \omega-3)N^2-4 (\bar \omega-1)^2 (\omega-1)) \, \times \nn \\
&&\times (\tr(13) \, \tr(2\bar 3) + \tr(1\bar 3) \, \tr(23))  \nn \\
&+& N^2((2\bar \omega \omega+3+3 \bar \omega^2 \omega^2)N^2 \, + \nn \\
&&  -2(\bar \omega-1)(4 \omega^2 \bar \omega+
 \omega^2 -\bar \omega \omega-\bar \omega+ \omega-4))
 (\tr(121 \bar 1)  + \tr(212 \bar 2)) \, + \nn \\
&-&{1 \over 3} N (6 (\bar \omega \omega+1)^2 N^4
+28(\bar \omega-1)^2 (\omega-1)^2  \, + \nn \\
&&-( \omega-1)(25 \omega \bar \omega^2-25 -29 \bar \omega \omega+
29 \bar \omega+2 \bar \omega^2-2  \omega) N^2 ) \, \times \nn \\
&& \times ( \tr(12) \, \tr(1\bar 1) +  \tr(21) \,  \tr(2\bar 2)) \, + \nn \\
&+& {4 \over 3} N (3(\bar \omega \omega+1)^2 N^4 -(\bar \omega-1)
(\omega-1)( 17 (\bar \omega \omega+ 1) +\omega+ \bar \omega) N^2 \, + \nn \\
&& +20 (\bar \omega-1)^2 ( \omega-1)^2) \ \tr(12) \, \tr(3\bar 3) \, .
\ea

\subsection{Finite corrections to the invariant
trilinear couplings at order $g^4$}

As already noted, most of the operators with vanishing
order $g^2$ and $g^4$ anomalous dimension receive finite
corrections to their 2-point functions at order $g^4$.
In principle the finite correction to the normalization
of the 2-point function of an operator is not in conflict with
conformal invariance or the vanishing of its anomalous dimension
at next perturbative orders. Moreover, if these operators are not
in the same supermultiplet with the generators of some
symmetry, their normalization is not protected, so we can
change it arbitrarily by a finite $g$ dependent
rescaling. There are, however, quantities which are invariant
under such a rescaling, namely the ratio of the
normalization of a 3-point function,
$\langle{\cal O}_1 {\cal O}_2 {\cal O}_3 \rangle$, to the
square root of the product of the normalizations of the
2-point functions, \ie
\be
C_{{\cal O}_1 {\cal O}_2 {\cal O}_3}  =
{\langle{\cal O}_1 {\cal O}_2 {\cal O}_3 \rangle \over
\sqrt{\langle{\cal O}_1 {\cal O}_1^{\dagger} \rangle
 \langle{\cal O}_2 {\cal O}_2^{\dagger} \rangle
\langle{\cal O}_3 {\cal O}_3^{\dagger} \rangle}} \, .
\label{trilinv}
\ee
Hence the real question is whether also these invariant
trilinear couplings  receive finite corrections or not.
The explicit calculation to order $g^4$ of the 3-point function
\be
\langle {\cal O}^{1112}(x_1) \ {\cal C}_{12}^{\dagger}(x_2) \
{\cal C}_{11}^{\dagger}(x_3) \rangle
\ee
(see eqs.~(\ref{d2chiral}) and~(\ref{O1112}) for the definition of the operators)
and of the associated 2-point functions, shows that such invariant
trilinear couplings get indeed a finite correction to this order.
Thus, the finite order $g^4$ corrections cannot be absorbed by a redefinition of
the operators.

For the reader's convenience we report below the ratios of the
order $g^4$ corrections to the tree-level expressions for the
relevant 2- and 3- point functions and for the invariant trilinear
coupling. For simplicity we have omitted a common numerical factor proportional
to $\zeta(3)$, but we kept the complete dependence on $\omega$ and $N$.
One gets
\ba
&& {\cal C}^{11}\Big{|}_{\rm 2pt}
:  {N^2 (N^2-4)(N^2((\omega\bar\omega+1)^2 -
2(\omega^2+\bar \omega^2)) + 2( \omega-1)^2(\bar \omega-1)^2)
\over 4 ((N^2-2)(\omega\bar\omega+1)+2(\omega+\bar\omega))^2}  ,
 \nn \\ { } \nn \\
&& {\cal C}^{12} \Big{|}_{\rm 2pt}  :  {N^2 (N^2-4)
( N^2(\omega \bar\omega-1)^2+2(\omega-1)^2(\bar\omega-1)^2)
\over 4 ((N^2-2)(\omega \bar \omega+1)+2(\omega+\bar\omega))^2} \, ,
 {} \nn \\
&& {\cal O}^{1112} \Big{|}_{\rm 2pt}
 \ : \  {(N^2-4) (\omega-1)(\bar\omega-1)
\over 2 (N^2-8) ((N^2-2)(\omega \bar \omega+1)+2(\omega+\bar\omega))^2}
\times  \nn \\
&& \qquad \qquad \quad \times [ (N^6-4 N^4-56 N^2+48)(\omega+1)(\bar\omega+1)  \, + \nn \\
&& \qquad \qquad \qquad -4(N^4-16N^2+24)(\omega+\bar\omega)]   \, ,
\label{2ptg4}
\ea
and
\ba
&& {\cal O}^{1112} {\cal C}_{12}^{\dagger}
{\cal C}_{11}^{\dagger} \Big{|}_{\rm 3pt}  \ : \
-{ (N^2-4)(\omega-1)(\bar\omega-1) \over ((N^2-2)
(\omega \bar \omega+1)+2(\omega+\bar\omega))^2 }
\times \nn \\
&& \qquad  \times ((N^4+6N^2-8)(\omega \bar\omega+1) +
(N^4+2 N^2+8)(\omega+\bar\omega))    \, , \nn
\\
 {} \nn \\
&&C_{{\cal O}^{1112} {\cal C}_{12}^{\dagger}
{\cal C}_{11}^{\dagger}} \ : \
-{ (N^2-4)(\omega-1)(\bar\omega-1) \over  (N^2-8)
((N^2-2)(\omega \bar \omega+1)+2(\omega+\bar\omega))^2 }
\times \nn \\
&& \qquad  \times ( N^2(3N^2-20)(\omega+1)(\bar\omega+1)
+40(\omega-1)(\bar\omega-1))  \, .
\label{3pinv}
\ea
The first two expressions in~(\ref{2ptg4}) for $|\omega|=1$ agree
with the results
of~\cite{PSZ}. Note, however, that the conclusion that the correction
to the 2-point function of ${\cal C}_{12}$ is suppressed in the planar
limit, while the correction to the 2-point function of ${\cal C}_{11}$
is not, crucially depends on the  choice $|\omega|=1$ made in~\cite{PSZ}.
For generic $\omega$ both corrections are non-vanishing in the large $N$ limit.
The invariant trilinear couplings
$C_{{\cal O}^{1111} {\cal C}_{11}^{\dagger}
{\cal C}_{11}^{\dagger}}$ of the two
operators  in~(\ref{d41111}) also receive
finite corrections  at order $g^4$.

\section{Conclusions}\label{Conclusions}

Our explicit analysis suggests that although the deformed theory
has only $\cN=1$ supersymmetry, it inherits many of the interesting
properties of $\cN=4$ SYM. It can be made (at least up to order $g^6$)
perturbatively finite. It is endowed with a rich spectrum of composite
operators with vanishing anomalous dimensions and a complicated operator
mixing pattern. It also shows some surprising new features, like the
finite perturbative corrections to the 2- and 3-point functions of
the operators with vanishing anomalous dimension. Finally we observe
that even in the large $N$ limit, the deformed and the undeformed
theories are significantly different.

A crucial but still open question is whether the deformed theory is
exactly conformal.

\section*{Acknowledgements}

It is a pleasure to thank Dan Freedman for extensive discussion and for
showing us the manuscript of ref.~\cite{FG} before publication.
The work of G.C.R. and Ya.S. was supported in part by INFN,
MIUR-COFIN contract 2003-023852,
EU contracts MRTN-CT-2004-503369 and MRTN-CT-2004-512194, INTAS
contract 03-51-6346, and NATO grant PST.CLG.978785.
E.S. is grateful to the organizers for the invitation to the conference
``The Legacy of Supergravity", Frascati, June 2005, were part of this work was done.
G.C.R thanks the Alexander von Humboldt Foundation for partial financial
support.

\end{document}